\documentclass[11pt,twoside]{article}

\usepackage{asp2006}
\usepackage{epsf}
\usepackage{psfig}
\usepackage{lscape}

\def\Figure#1#2[#3] {
\centerline{
\scalebox{#3}{
\includegraphics{#1.#2}
}}}

\begin{document}
\title{ The Impact of LSST on Asymptotic Giant Branch Star Research }
\author{ \v{Z}eljko Ivezi\'{c}, for the LSST Collaboration}
\affil{ Department of Astronomy, University of Washington, Seattle, WA 98155,
  USA}
 
\begin{abstract}
The Large Synoptic Survey Telescope (LSST) is currently by far the 
most ambitious proposed ground-based optical survey. With initial 
funding from the US National Science Foundation (NSF), Department of 
Energy (DOE) laboratories and private sponsors, the design and 
development efforts are well underway at many institutions, including 
top universities and leading national laboratories. The main science 
themes that drive the LSST system design are Dark Energy and Matter, 
the Solar System Inventory, Transient Optical Sky and the Milky Way 
Mapping. The LSST system, with its 8.4m telescope and 3,200 Megapixel
camera, will be sited at Cerro Pachon in northern Chile, with the first
light scheduled for 2013. In a continuous observing campaign, LSST will 
cover the entire available sky every three nights in two photometric bands 
to a depth of V=25 per visit (two 15 second exposures), with exquisitely 
accurate astrometry and photometry. Over the proposed survey lifetime of 
10 years, each sky location would be observed about 1000 times, with the
total exposure time of 8 hours distributed over six broad photometric 
bandpasses (ugrizY).  This campaign will open a movie-like window on 
objects that change brightness, or move, on timescales ranging from 10
seconds to 10 years. The survey will have a data rate of about 30 TB/night,
and will collect over 60 PB of data over its lifetime, resulting 
in an incredibly rich and extensive public archive that will be a treasure 
trove for breakthroughs in many areas of astronomy. I describe how 
this archive will impact the AGB star research and speculate how the 
system could be further optimized by utilizing narrow-band TiO and CN
filters. 
\end{abstract}

\section{ The Large Synoptic Survey Telescope }

Three recent committees comissioned by the US National Academy of 
Sciences\footnote{
  Astronomy and Astrophysics in the New Millennium, NAS 2001; 
  Connecting Quarks with the Cosmos: Eleven Science Questions for the New Century, NAS 2003; 
  New Frontiers in the Solar System: An Integrated Exploration Strategy, NAS 2003.   
}
concluded that a dedicated wide-field imaging telescope with an effective aperture 
of 6--8 meters is a high priority for US planetary science, astronomy, and physics 
over the next decade. The LSST system (Tyson 2002) includes such a telescope, and
is designed to obtain sequential images covering the entire visible sky every
few nights. Detailed simulations that include measured weather statistics and
a variety of other effects which affect observations predict that each sky
location can be visited about 100 times per year, with two 15 second long
exposures per visit.  

The range of scientific investigations which would be enabled by such a 
dramatic improvement in survey capability is extremely broad. The main
science themes that drive the LSST system design are
\begin{enumerate}
\item {\bf Constraining Dark Energy and Dark Matter} using a variety of 
           probes and techniques whose synergy will fudamentally test our 
           cosmological assumptions and gravity theories 
\item {\bf Taking an Inventory of the Solar System} and extending the boundaries
           of our reach in distance and detectable size of potentially hazardous
           asteroids 
\item {\bf Exploring the Transient Optical Sky} by characterizing known classes 
           of object and discovering new ones
\item {\bf Mapping the Milky Way} all the way to its edge with high-fidelity
\end{enumerate}

\subsection{     The LSST  Reference Design    }

The LSST reference design\footnote{More details about LSST system are available at 
http://www.lsst.org.}, with an 8.4 m diameter primary mirror, standard
filters, and current detector performance, reaches a depth equivalent to 
$V$$\sim$25 in 30 seconds\footnote{An LSST exposure
time calculator has been developed and is publicly available at 
http://tau.physics.ucdavis.edu/etc/servlets/LsstEtc.html.} over its entire 10 square
degree field. The key figure of merit for a large survey telescope is the
\'{e}tendue - the product of the collecting area of its primary mirror and the
field of view  ($A\Omega$). The solid angle of sky that can be surveyed to a
limiting depth per unit time is proportional to the \'{e}tendue. LSST will
provide an order of magnitude
larger \'{e}tendue than any existing facility, and at least a factor five
larger \'{e}tendue than any other planned or proposed facility.  A unique
consequence of this very high \'{e}tendue is that many different science
programs can utilize a common observing strategy, yielding a single common
database. One can think of this as {\it massively parallel astrophysics}:
Rather than devoting a distinct set of observations to each area of science, 
a single universal set of observations feeds all science investigations in parallel.

\begin{figure}[!t]
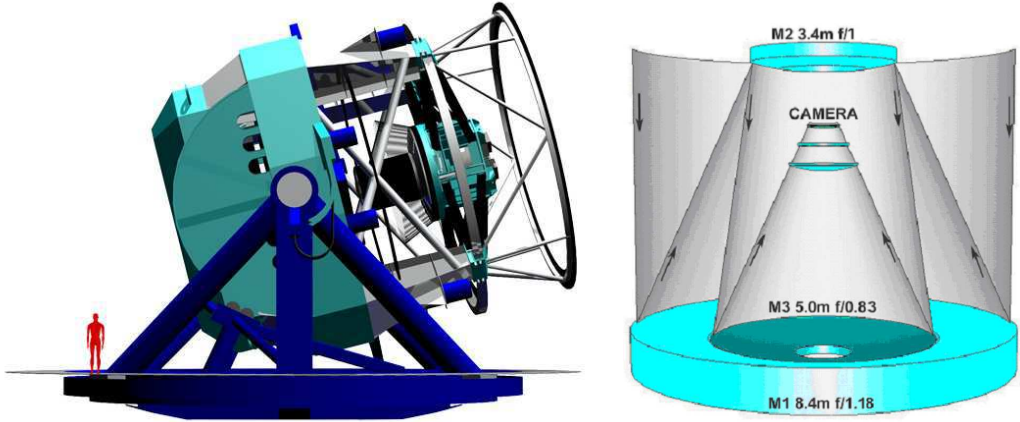

\phantom{x}
\vskip -2.0in
\hskip -1.1in 
\Figure{telescope2S}{ps}[0.55]
\vskip -4.7in
\hskip 1.6in
\Figure{opticsS}{ps}[0.3]
\vskip -0.7in
\caption{The left panel shows baseline design for LSST telescope, current as of 
April 2006. The telescope will have an 8.4-meter primary mirror, and a 
10-square-degree field of view. The right panel shows LSST baseline optical design 
with its unique monolithic mirror: the primary and tertiary mirrors are coplanar 
and their surfaces will be polished into single substrate.}
\label{tel}
\end{figure}

\begin{figure}[ht]
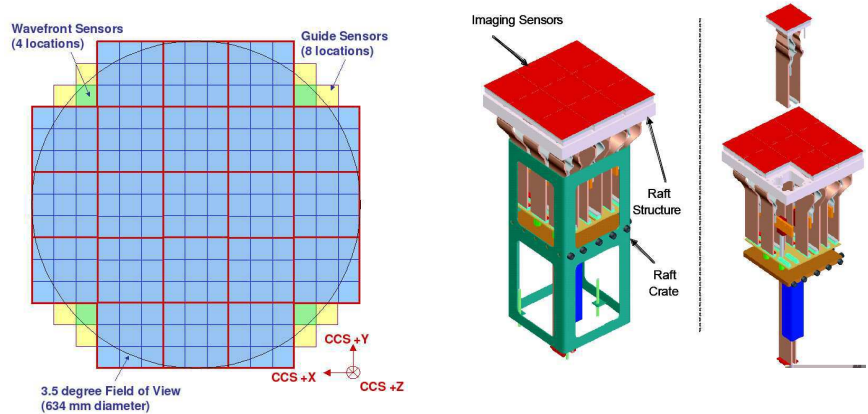

\vskip -0.2in
\hskip -1.1in 
\Figure{fovS}{ps}[0.3]
\vskip -3.5in
\hskip 1.3in
\Figure{raftS}{ps}[0.3]
\vskip -0.5in
\caption{The left panel shows LSST focal plane. Each cyan square represents one 
4096x4096 pixel large sensor. Nine sensors are assembled together in a raft. There 
are 189 science sensors, each with 16.8 Mpix, for the total pixel count of 3.2 Gpix.
LSST raft module with integrated front-end electronics and thermal connections is
shown in the right panel.}
\label{fov}
\end{figure}

This large etendue is achieved in a novel 
three-mirror design (modified Paul-Baker) with a very fast f/1.2 beam, together with 
a 3200 megapixel camera with state-of-the-art detectors. The baseline designs
for telescope and camera are shown in Figs.~\ref{tel}--\ref{fov}, and the
main system parameters are summarized in Table~1.

The LSST survey will open a movie-like window on objects that change
brightness, or move, on timescales ranging from 10 seconds to 10 years.
The survey will have a data rate of about 30 TB/night (more than one
complete Sloan Digital Sky Survey per night), and will collect over 60 PB
of data over its lifetime, resulting in an incredibly rich and extensive
public archive that will be a treasure trove for breakthroughs in many areas 
of astronomy. In the next section I describe how this archive will impact the 
AGB star research and speculate how the LSST system could be further optimized by 
utilizing narrow-band TiO and CN filters.

\begin{table}[!b]
\centerline{
\begin{tabular}{|l|l|}
\hline 
   Quantity                 &     Baseline Design Specification    \\
\hline  
Optical/mount Configuration &  3-mirror modified Paul-Baker; alt-azimuth          \\
Final f-Ratio, aperture              &  f/1.25, 8.4 m                \\
Field of view area, \'etendue        &  9.6 deg$^2$,   318 m$^2$deg$^2$     \\
Plate Scale, pixel count     &  50.9 $\mu$m/arcsec (0.2'' pix), 3.2 Gigapix  \\
Wavelength Coverage, filters &  320 -- 1050 nm, ugrizY                        \\
Single visit depths (5$\sigma$) & $u:23.9$, $g:25.0$, $r:24.7$, $i:24.0$, $z:23.3$, $Y:22.1$ \\
Mean number of visits      &  $u:70$, $g:100$, $r:230$, $i:230$, $z:200$, $Y:200$ \\ 
Final (coadded) depths (5$\sigma$) & $u:26.3$, $g:27.5$, $r:27.7$, $i:27.0$, $z:26.2$, $Y:24.9$ \\
\hline                         
\end{tabular}                  
}
\caption{The LSST Baseline Design and Survey Parameters}
\end{table}             

\section{ The Impact of LSST on AGB Star Research }

The LSST survey will yield continuous overlapping images of 20,000 square
degrees of sky in six broad-band filters covering the wavelength range 320 --
1050 nm. Each sky location will be observed about 1000 times (total exposure 
time of 8 hours) over the survey lifetime of 10 years. These data can be 
utilized for finding AGB stars by using several methods:
\begin{enumerate}
\item {\bf Optical identifications of IR counterparts} For example, if 
IRC$+$10216 were at 40 kpc, it would have $i$=27, $z$=25 and $Y$=23 (based on 
SDSS observations), which is brighter than LSST faint limits in these bands. 
Therefore, even stars with exceedingly thick dust shells and barely detected 
by IRAS will be detectable in the $i$, $z$, and $Y$ bands by LSST throughout
the Galaxy.
\item {\bf Search for spatially resolved envelopes} As demonstrated by SDSS \
observations of IRC$+$10216, LSST will be able to detect and {\it resolve} an
IRC$+$10216-like envelope at a distance of 15 kpc!
\item {\bf Color selection} Extremely red colors of dusty AGB stars are a very 
      distinctive characteristic; color-selected LSST samples will be able to 
      trace structure throughout the Local Group and beyond.
\item {\bf Variability} Large optical amplitudes and $\sim$1000 observations 
      over 10 years will be a powerful detection method for AGB stars (e.g. 
      LSST will detect over a hundred million variable stars).
\end{enumerate}

It is evident that LSST, although driven by different science goals,
will be a powerful machine for discovering and characterizing AGB stars. 
This ability could be further enhanced by utilizing narrow-band filters.

\subsection{ Specialized narrow-band filters }

\begin{figure}[!t]
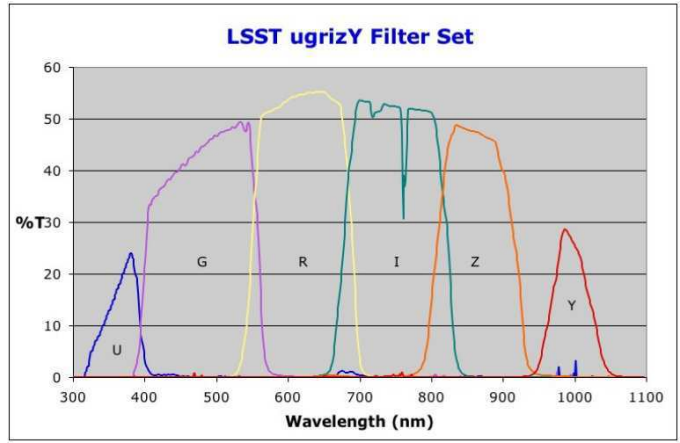

\vskip -1.4in
\Figure{lsstfiltersS}{ps}[0.5]
\vskip -1.9in
\caption{The current design of the LSST bandpasses. The y axis shows the
overall system throughput. The bandpasses are similar to those employed by
the Sloan Digital Sky Survey, with addition of the Y band.}
\label{filters}
\end{figure}

The current LSST baseline design includes six broad-band filters. The system 
throughput as a function of wavelength for these bandpasses is shown in 
Fig.~\ref{filters}. The ability of LSST to characterize AGB stars (e.g.
C vs. O type classification) could be further enhanced by adding narrow-band 
filters. For example,
the so-called TiO (7780~$\AA$) and CN (8120~$\AA$) filters introduced by 
Wing (1971) have been successfully used by a number of groups (Cook, Aaronson 
\& Norris 1986; Kerschbaum et al. 2004, Battinelli \& Demers 2005, and
references therein) for identification and characterization of late-type 
stars. 

The LSST Scientific Requirements Document allows for about 10\% of
the observing time (300 nights) to be allocated to specialized programs. 
If only 2 nights ($<$0.1\% of the total observing time) were allocated 
to a narrow-band survey, it would be possible to cover about 10,000
deg$^2$ of sky in each band. Such a time allocation would match the
cost of procuring filters to the cost of LSST system (about 150,000 USD
per filter and per observing night). 

Assuming 100 $\AA$ wide filters, the faint limits would be about $m=22-22.5$
(AB). This is about 0.5--1 mag shallower than e.g. a recent study of And II \
by Kerschbaum et al. (2004), but the surveyed area would be over 1,000,000 
times larger! Furthermore, it is noteworthy that the deep and exceedingly 
accurate broad-band photometry will come for ``free'', and will include many 
epochs which can be used to reject foreground Galactic M dwarfs by 
the lack of variability. 

This program may represent an exciting opportunity for the AGB star
community. In order to execute such a program, this community may wish
to organize a working group which would have three main goals: fundraising 
for the filter procurement, securing the allocation of observing time  
from the LSST Collaboration, and the timely analysis of narrow-band
survey data.

\acknowledgements
I acknowledge my numerous LSST collaborators for their efforts in the
design and development of LSST system, including the figures used in this 
contribution. I am thankful to Jill Knapp for inspiration that in part
motivated this work.

\end{document}